\documentclass[prb,twocolumn,showpacs,preprintnumbers,amsmath,amssymb]{revtex4-1}

\usepackage{graphicx}
\usepackage{dcolumn}
\usepackage{bm}

\begin{document}

\title{Electronic Structure and Fermiology of Superconducting LaNiGa$_2$}

\author{David J. Singh}

\affiliation{Materials Science and Technology Division,
Oak Ridge National Laboratory, Oak Ridge, Tennessee 37831-6056}

\date{\today}

\begin{abstract}
We report electronic structure calculations for the layered centrosymmetric
superconductor LaNiGa$_2$, which has been identified as having
a possible triplet state based on evidence for time reversal symmetry
breaking.
The Fermi surface has several large sheets and is only moderately
anisotropic, so that the material is best described as a three dimensional
metal.
These include sections that are open in the in-plane direction as well
as a section that approaches the zone center.
The density of states is high and primarily derived from Ga $p$ states,
which hybridize with Ni $d$ states.
Comparing with experimental specific heat data, we infer a superconducting
$\lambda \le$ 0.55, which implies that this is a weak to intermediate coupling
material.
However, the Ni occurs in a nominal
$d^{10}$ configuration in this material,
which places the compound far from magnetism.
Implications of these results for superconductivity are discussed.
\end{abstract}


\maketitle

\section{introduction}

Hillier and co-workers recently discovered the appearance of spontaneous
magnetic fields with onset at the superconducting critical temperature
in samples of the centrosymmetric intermetallic compound
LaNiGa$_2$ using muon spin rotation ($\mu$SR).
\cite{hillier}
Symmetry analysis implies that LaNiGa$_2$, which is
a $\sim$2 K superconductor, \cite{zeng,aoki}
is a triplet superconductor with a non-unitary state.
\cite{hillier}
One mechanism for obtaining triplet superconductivity is nearness to
ferromagnetism as in the likely triplet superconductor
Sr$_2$RuO$_4$.
\cite{mackenzie,rice,mazin}
Interestingly, Ni is a ferromagnet and intermetallic Ni$_3$Ga
is a highly renormalized itinerant paramagnet near ferromagnetism.
\cite{hayden,aguayo}

There is, however, little other data available about the superconducting
properties of LaNiGa$_2$.
So far, three reports are all based on polycrystalline samples prepared
by arc melting using different source material. Aoki and co-workers
reported bulk superconductivity with $T_c$=2.01 K
(onset at 2.1 K) on a sample with
a residual resistivity ratio of 34 and residual resistivity of $\sim$1.5
$\mu\Omega$ cm, while Zeng and co-workers obtained $T_c$=1.97 K,
on a sample with a residual resistivity ratio of 5.2 and
residual resistivity of 14.1 $\mu\Omega$ cm.

The purpose of this paper is to report the electronic structure and
related properties in relation to the superconductivity of this material.
Our density functional calculations
were based on
the generalized gradient approximation of Perdew, Burke and Ernzerhof,
\cite{pbe}
and used the general potential linearized augmented planewave (LAPW) method,
\cite{singh-book} as implemented in the WIEN2k code. \cite{wien2k}
The LAPW sphere radii employed were 2.5 bohr, 2.2 bohr and 2.0 bohr
for La, Ni and Ga, respectively.
Relativity was included at the scalar relativistic level for the valence 
states (the core states were treated fully relativistically).
We used highly converged basis sets corresponding to
$R_{min}k_{max}$=9.0, where $k_{max}$ is the interstitial planewave
cut-off and $R_{min}$=2.0 bohr is the smallest sphere radius,
as well as dense Brillouin zone samples, i.e. a
32x32x32 mesh for the calculations of the Fermiology and
a 16x16x16 mesh for the fixed spin moment calculations.
The semi-core states (La 5$s$, 5$p$, Ni 3$p$ and Ga 3$d$) were included
with the valence electrons using local orbitals.
We used the standard LAPW basis, as opposed to the so-called APW+lo
basis. \cite{sjo}

\begin{figure}
\includegraphics[width=\columnwidth]{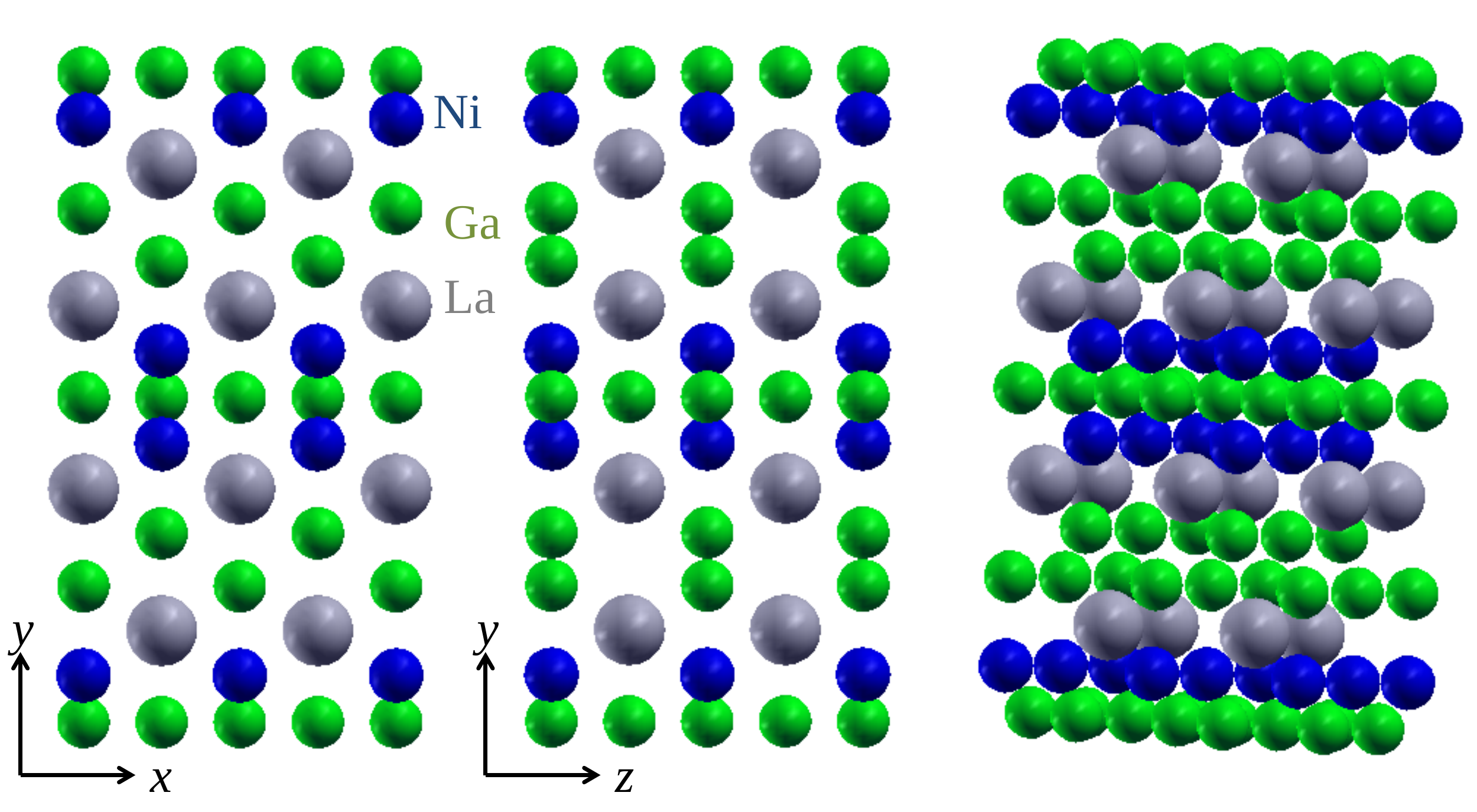}
\caption{(color online) Crystal structure of LaNiGa$_2$ showing
the coordinate system used here. The structure depicted
is based on the experimental lattice parameters with relaxed internal
coordinates.}
\label{struct}
\end{figure}

LaNiGa$_2$ occurs in an orthorhombic structure, with space group, \#65, $Cmmm$,
and two formula units per primitive cell.
\cite{romaka}
The calculations were done using the experimental lattice parameters,
$a$=4.29 \AA, $b$=17.83 \AA{} and $c$=4.273 \AA, \cite{romaka}
with internal atomic
coordinates determined by total energy minimization.
The structure is depicted in Fig. \ref{struct} and the
calculated atomic coordinates are given in Table \ref{struct-tab}.
As may be seen, the structure is layered along the $b$-axis. This
might suggest an effectively two dimensional electronic structure, but this
is not what we find (see below).

\begin{figure}
\includegraphics[width=\columnwidth]{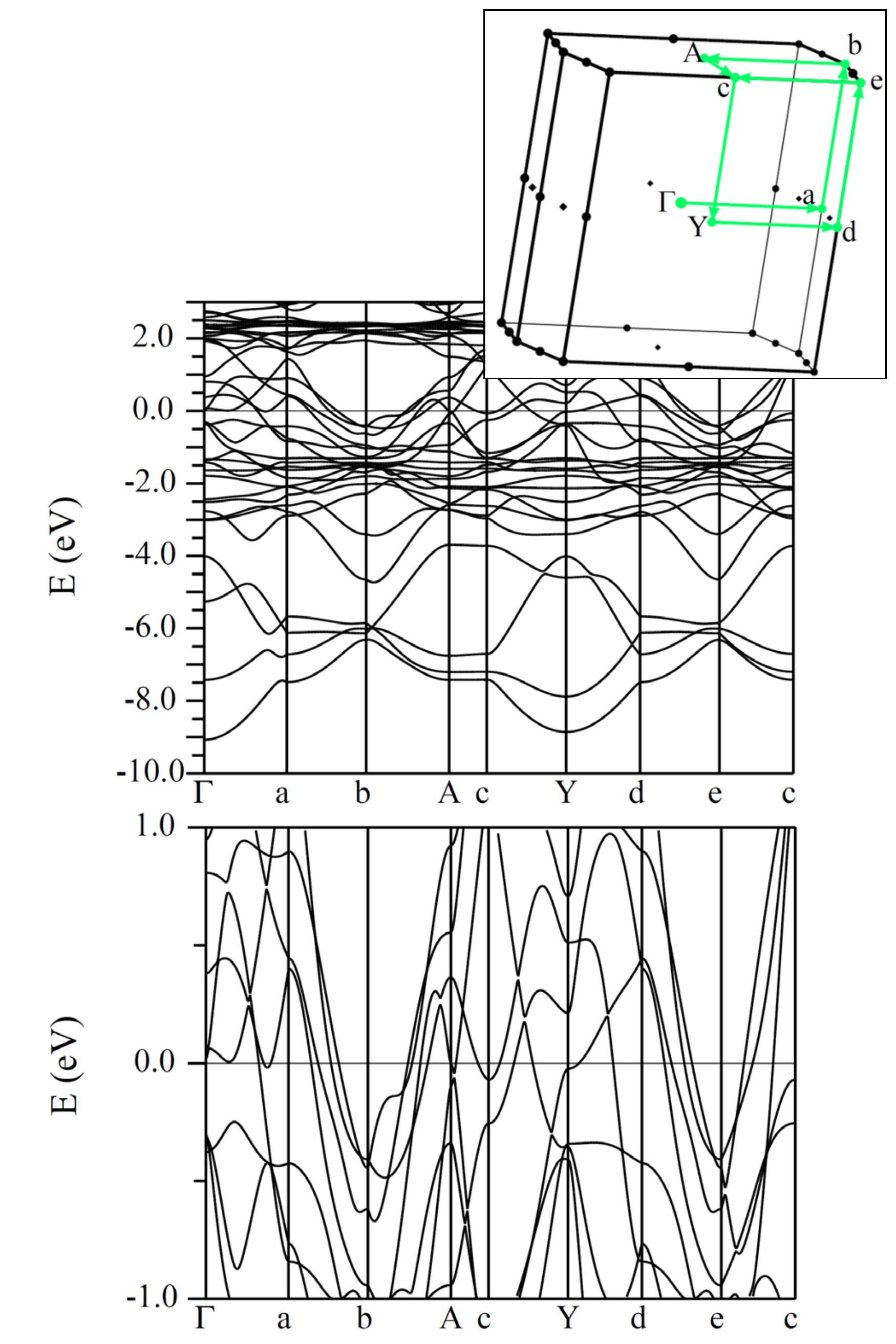}
\caption{(color online) Density functional band structure of LaNiGa$_2$
as obtained for the relaxed crystal structure.
The dotted horizontal lines at 0 eV denote the Fermi energy, $E_F$.
The lower panel is a blow-up around $E_F$.
The path through the zone and labels are shown in the inset.}
\label{bands}
\end{figure}

\begin{figure}
\includegraphics[width=\columnwidth,angle=0]{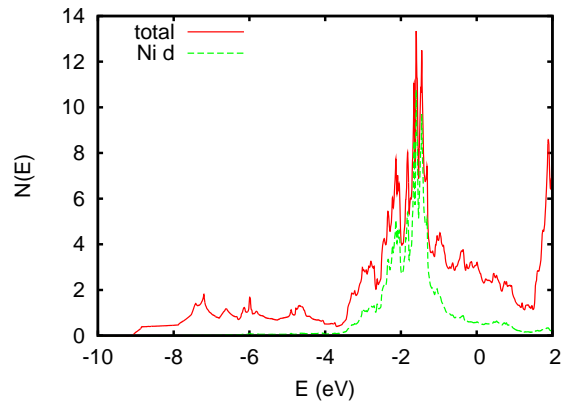}
\caption{(color online) Electronic density of states and $d$ projection
onto the Ni LAPW sphere on a per formula unit basis.}
\label{dos}
\end{figure}

We start with the large energy
scale features of the band structure and density of states (DOS), which are
shown in Figs. \ref{bands} and \ref{dos}, respectively.
The band structure shows four bands in the energy range from -9 eV to -4 eV
(all energies are given with respect to the Fermi energy $E_F$).
These are derived primarily from the Ga $s$ orbitals (note that there
are two formula units per primitive unit cell, i.e. 4 Ga atoms).
The unoccupied flat bands starting at $\sim$2 eV are the La $4f$
states.

Between the Ga $s$ bands and the La $f$ resonance there are 
dispersive bands of primarily Ga $p$ character and additional flatter
bands centered at $\sim$-2 eV. These occupied flat bands are the Ni $d$
bands, which mix with the Ga $p$ bands in the energy range around -2 eV.
This is clearly seen in the DOS, which has a prominent peak of 
Ni $d$ character centered near -2 eV, with a width of $\sim$ 2 eV.
While one may observe that there is some Ni $d$ character at and
above $E_F$, this is a minor component that arises because of hybridization
in the Ga $p$ derived bands.
This means that the Ni $d$ bands are nominally occupied in this compound,
and correspondingly that Ni occurs in a $d^{10}$ configuration.

\begin{table}
\caption{Internal atomic coordinates of $Cmmm$ LaNiGa$_2$ as
determined by total energy minimization. The coordinates
are with respect to the experimental lattice parameters,
$a$=4.29 \AA, $b$=17.83 \AA{} and $c$=4.273 \AA}
\begin{tabular}{lccc}
\hline
~~~~~~~~~ & ~~~~$x$~~~~ & ~~~~$y$~~~~ & ~~~~$z$~~~~ \\
\hline
La (4j) & 0.0 & 0.3591 & 0.5 \\
Ni (4i) & 0.0 & 0.0719 & 0.0 \\
Ga1 (4i) & 0.0 & 0.2092 & 0.0 \\
Ga2 (2d) & 0.0 & 0.0 & 0.5 \\
Ga3 (2b) & 0.0 & 0.0 & 0.0 \\
\hline
\end{tabular}
\label{struct-tab}
\end{table}

The implication is that electronic structure near the Fermi energy
in LaNiGa$_2$ is derived from $sp$ bands of primarily Ga $p$ character,
hybridized with Ni $d$ states. This is surprising for an unconventional
superconductor, where one might naturally suppose that triplet pairing
is a consequence of magnetism associated with the transition element
or perhaps other correlation effects due to an open $d$ or $f$ shell.

Turning to the low energy properties, there are several bands
crossing the $E_F$ as shown in Fig. \ref{bands}.
We obtain $N(E_F)$=3.19 eV$^{-1}$ on per formula unit both spins basis,
which corresponds to a bare specific heat coefficient,
$\gamma_{bare}$=7.52.
Zeng and co-workers \cite{zeng}
reported a specific heat coefficient, 
$\gamma$=11.64 mJ/mol K$^2$,
which implies an enhancement, $\gamma$=$\gamma_{bare}(1+\lambda)$,
with $\lambda$=0.55.
This is consistent with the conclusion of Zeng and co-workers
that LaNiGa$_2$ is a weakly coupled superconductor.

\begin{figure}
\includegraphics[width=\columnwidth,angle=0]{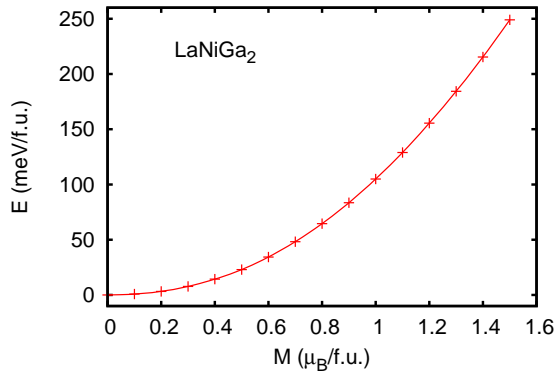}
\caption{(color online)
Energy as a function of constrained spin magnetization from fixed
spin moment calculations. The energy and magnetization are on a 
per formula unit basis, and the energy is relative to the
non-spin-polarized case. The symbols are calculated points, while
the curve is an interpolation.}
\label{fsm}
\end{figure}

\begin{figure}
\includegraphics[width=\columnwidth]{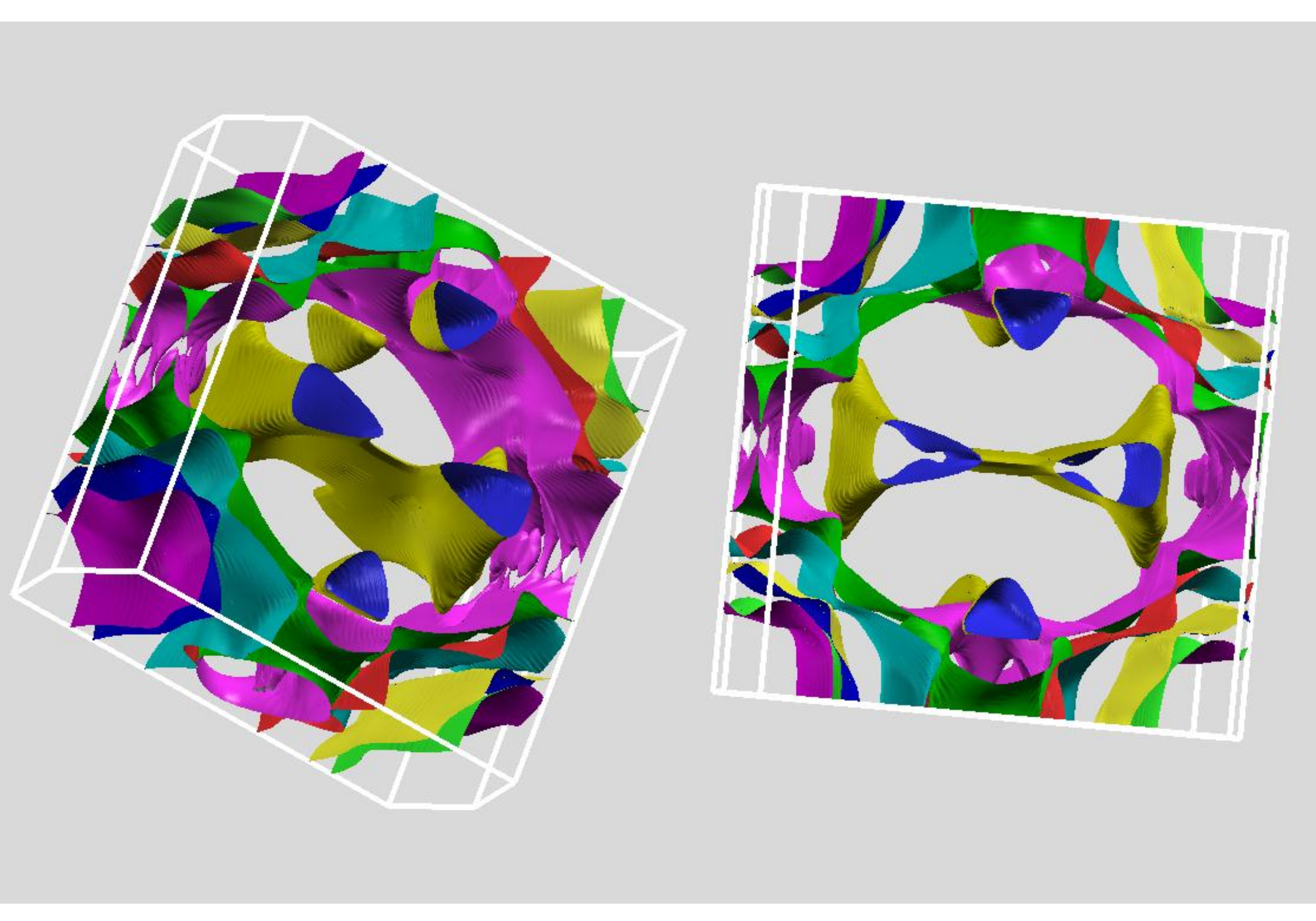}
\caption{(color online) Two views of the calculated Fermi surface of LaNiGa$_2$.
$\Gamma$ is at the center of each plot.
The zone and labels are given in the inset of Fig. \ref{bands}. The coloring
is arbitrary and is used to distinguish the different sections.}
\label{fermi}
\end{figure}

The substantial value of $N(E_F)$ would imply that the material is either
an itinerant ferromagnet or close to it if the bands near the Fermi energy
were primarily Ni derived.
However, this is not the case and the Ni $d$ component
of the density of state is not large, having a value of 0.59 eV$^{-1}$
per formula unit both spins. Taking a typical Ni Stoner $I$ of 1 eV,
\cite{janak}
this yields $NI\sim$0.3 (note that
the $N(E_F)$ in the Stoner formula is per spin).
This is far less than unity, indicating that
this material is not near magnetism.
We did fixed spin moment calculations to confirm this. The energy
as a function of constrained moment is shown in Fig. \ref{fsm}.
As may be seen there is no indication of metamagnetism or nearness
to a ferromagnetic state.

The calculated Drude plasma energies are
$\hbar\Omega_{p,xx}$=4.40 eV,
$\hbar\Omega_{p,yy}$=2.11 eV, and
$\hbar\Omega_{p,zz}$=4.71 eV.
Within Boltzmann transport theory, the conductivity is related to the plasma
frequency,
$\sigma_{xx}\propto N(E_F)<v_x^2>\tau\propto\Omega_{p,xx}^2\tau$, and
similarly for the other directions, where $\tau$ is an inverse scattering
rate. Therefore the transport is predicted to be three dimensional,
and only moderately 
anisotropic, with the $b$-axis conductivity lower than the in-plane
conductivity by a factor of $\sim$5.
This three dimensionality is perhaps not surprising in light of the
fact that the electronic structure near $E_F$ is derived from bands
that have primary Ga $p$ character, hybridized with Ni $d$ states, rather
than being mainly derived from the more compact Ni $d$ orbitals.

As mentioned, there are several bands that cross $E_F$. The
Fermi surface is shown in Fig. \ref{fermi}.
There are several large sheets, including sheets near the zone center as
well as the zone corners. Additionally, besides open sheets along $k_y$,
which is the direction perpendicular to the layers, there are open sheets
along both of the in-plane ($k_x$,$k_z$) directions as well.
Specific heat measurements
show an exponential dependence below $T_c$ (Ref. \onlinecite{zeng}),
which indicates a fully gapped superconducting state.
In a triplet superconductor the order parameter must change
sign under inversion through the $\Gamma$ point.
In this context the combination of a fully gapped state and the complex
open Fermi surfaces in all crystallographic directions and sheets
very close to the zone center is unexpected since
simple triplet states would not be fully gapped on such a Fermi surface.

The thermopower of a metal is sensitive to the details of
the band structure at the Fermi
energy. We calculated the thermopower within the constant
scattering time approximation based on the first principles band
structure. We used the BoltzTraP code for this purpose. \cite{boltztrap}
We obtain negative values of 
$S_{xx}$(300 K)= -8.3 $\mu$V/K,
$S_{yy}$(300 K)= -9.6 $\mu$V/K, and
$S_{zz}$(300 K)= -0.3 $\mu$V/K.
Averaging these values with the conductivity, we obtain
$S_{av}=(S_{xx}\sigma_{xx}+S_{yy}\sigma_{yy}+S_{zz}\sigma_{zz})/
(\sigma_{xx}+\sigma_{yy}+\sigma_{zz})=$
-4.6 $\mu$V/K at 300 K, which is very close to the value
of $\sim$-5 $\mu$V/K from Fig. 8 of Ref. \onlinecite{aoki}.
This provides support for the calculated Fermi surface.

To summarize the results of the calculations, we find
Ni to be in a nominal $d^{10}$ state. LaNiGa$_2$ has
a complex three dimensional Fermi surface, derived mainly from $sp$ states,
that hybridize with Ni $d$ states. This Fermi surface includes open
sections in all three crystallographic directions and additionally has a section
near the zone center. We do not find proximity to ferromagnetism but we
do find a moderately high $N(E_F)$, which in conjunction with experimental
specific heat data suggests a modest $\lambda$ consistent with weak coupling.

The data raise some other questions about the superconductivity of LaNiGa$_2$.
First of all,
we do not find heavy bands. However, the weak dependence of
$T_c$ on residual resistivity \cite{aoki,zeng} is most readily explained in
a triplet scenario if the bands are very heavy (as in a heavy Fermion) so
that the coherence length becomes
very short. Actually, besides the dispersive bands,
we note that the coherence length $\zeta$=28 nm determined from the
$\mu$SR measurements is not so short although it is shorter
than the 66 nm coherence length of Sr$_2$RuO$_4$. \cite{mackenzie}
Secondly, there is a difficulty in identifying a plausible
pairing interaction.
While a purely attractive interaction, such as the electron phonon
interaction can be pairing for a triplet state provided that it has strong
momentum dependence, as is easily seen from the
gap equation it will be more pairing for a conventional singlet $s$-wave
state. Therefore even in this case an additional repulsive interaction
will be needed. Two possible interactions are the Coulomb repulsion, 
and spin-fluctuations. However, the dispersive $sp$ bands argue against these
in LaNiGa$_2$. Also the modest value of the $\lambda$ inferred from the
specific heat does not leave much room for competing interactions (note
that in a case where one has repulsive and attractive interactions they will
partially cancel for the superconducting $\lambda$ but will be additive for the
specific heat $\lambda$).

Nonetheless, it is a fact that time reversal symmetry breaking has been
observed by $\mu$SR at the bulk $T_c$ in samples of this material. \cite{hillier}
Further characterization of LaNiGa$_2$ and its superconducting properties
would be highly desirable, particularly using pure phase single crystals
if these can be made.

\acknowledgments

I am thankful for helpful discussions with
J. Quintanilla, J.F. Annett and A.D. Hillier.
This work was supported by the Department of Energy, Basic Energy Sciences,
Materials Sciences and Engineering Division.

\bibliography{LaNiGa2}

\end{document}